\documentclass{aastex}

\begin{document}

\title{VLBA observations of $z > 4$ radio-loud quasars}

\author{Emmanuel Momjian\altaffilmark{1}}
\affil{National Radio Astronomy Observatory, P. O. Box O, Socorro, NM, 87801}
\email{emomjian@nrao.edu}
\altaffiltext{1}{Current address: NAIC, Arecibo
Observatory, HC3 Box 53995, Arecibo, PR 00612, emomjian@naic.edu}

\author{Andreea O. Petric}
\affil{Astronomy Department, Columbia University, New York, NY, 10027}
\email{andreea@astro.columbia.edu}

\author{Christopher L. Carilli}
\affil{National Radio Astronomy Observatory, P. O. Box O, Socorro, NM, 87801}
\email{ccarilli@nrao.edu}

\begin{abstract}

We present high resolution ($\leq 20$~mas) observations of the radio
continuum emission at 1.4 GHz from three high-redshift quasars:
J1053$-$0016 ($z=4.29$), 1235$-$0003 ($z=4.69$), and J0913$+$5919
($z=5.11$), thereby doubling the number of $z>4$ radio-loud quasars
that have been imaged at mas resolution. The
observations were carried out with the Very Long Baseline Array
(VLBA) of the NRAO.  All three sources are unresolved in these observations,
with source size limits of a few mas. In all cases the flux densities
measured by the VLBA are within 10$\%$ of those measured with the VLA,
implying that the sources are not highly variable on yearly
timescales.  We find no indication for multiple images that might be
produced by strong gravitational lensing on scales from 20 mas (VLBA)
to a few arcseconds (VLA), to dynamic range limits of $\sim 100$.

\end{abstract}


\section{INTRODUCTION}

Surveys such as the Sloan Digital Sky Survey (SDSS) \citep{YOR00} and
the Digitized Palomar Sky Survey \citep{DJO99} have
revealed large samples of quasi-stellar objects (QSOs) out to $z \sim
6$. Studies by \citet{FAN02} have
shown that at such a high redshift we are approaching the epoch of
reionization, the edge of the ``dark ages'', when the first stars and
massive black holes were formed. Eddington-limit arguments suggest that
the supermassive black holes at the center of these QSOs are on the order of
$10^9~M{_\odot}$. If the correlation between bulge and black-hole masses \citep{GEB00,FEME00}
also holds at these high redshifts, then these sources have associated spheroids
with masses on the order of $\sim 10^{12}~M_\odot$. It is challenging to explain the formation
of such massive structures on relatively short timescales {($\sim$~1~Gyr)}.
However, \citet{WL02} estimate that almost one third of known quasars at $z\sim6$
ought to be lensed by galaxies along the line of sight. If these quasars are indeed
gravitationally lensed, the estimated masses of their associated spheroids could be smaller 
by up to an order of magnitude; this would allow a less 
efficient assembly process. 

High resolution radio observations of high-redshift radio-loud quasars
can be used to test for strong gravitational lensing by looking for
multiple imaging on scales from tens of milliarcseconds (mas) to
arcseconds. Also, Very Long Baseline Interferometry (VLBI) observations of core-jet radio sources in
quasars over a large range in redshift have been used to constrain the
cosmic geometry, under the assumption that such sources are (roughly)
'standard rulers' \citep{GKF99}. In general, the high resolution of the VLBI
observations permits a more detailed look at the physical structures in the most
distant cosmic sources.

To date, only three radio loud quasars at $z > 4$ have been imaged at
mas resolution \citep{FRE97,FRE03}. In this paper we
present Very Long Baseline Array (VLBA) observations of three more quasars at $z> 4$: SDSSp
J105320.42$-$001649.7 at $z=4.29$, SDSSp J123503.04$-$000331.7 at
$z=4.69$, and SDSSp J091316.56$+$591921.5 at $z=5.11$ (hereafter
J1053$-$0016, J1235$-$0003, and J0913$+$5919). J1053$-$0016
was first identified by \citet{IMH91} and \citet{STD94} as BRI 1050$-$0000,
while the other two sources were discovered in the
five-color imaging data from the SDSS \citep{FAN00,AND01}. All three
quasars are known to be radio-loud \citep{CAL01,PET03}

Throughout this paper, we assume a flat cosmological model with
$\Omega_{m}=0.3$, $\Omega_\Lambda=0.7$, and
${H_{0}=65}$~km~s$^{-1}$~Mpc$^{-1}$. In this model, 1~mas
corresponds to about 7~pc at $z=4.5$.

\section{OBSERVATIONS AND DATA REDUCTION}

The observations were carried out at 1425~MHz using the ten stations of
the Very Long Baseline Array (VLBA) of the NRAO\footnote{The National
Radio Astronomy Observatory is a facility of the National Science
Foundation operated under cooperative agreement by Associated
Universities, Inc.}. The sources J0913$+$5919, J1235$-$0003, and
J1053$-$0016, were observed on the 1st, 21st, and 29th of March 2003, respectively.
The total observing time on each source was 7~hr.

The observations employed nodding-style phase referencing with a cycle time of 3.5~minutes.
The phase calibrators of J1053$-$0016, J1235$-$0003, and J0913$+$5919, were
J1048$+$0055, J1232$-$0224 , and J0921$+$6215, respectively.
Phase-referencing observations allow the determination of the absolute position
of the target source and its components, if any, from the position of the calibrator
\citep {WAL99}.
In applying the phase-referencing technique, the accuracy of the calibrator position is
important. The positions of our phase-reference sources are obtained from the
VLBA Calibrator Survey \citep{BEA02}, and are accurate to about 1~mas.

Two adjacent 8~MHz baseband channel pairs were
used in the observations of each source, both with right and left-hand
circular polarizations, sampled at two bits. The data were correlated
at the VLBA correlator in Socorro, NM, with 2~s correlator integration
time. Data reduction and analysis were performed using NRAO's Astronomical Image
Processing System (AIPS).

After applying {\it a priori} flagging, amplitude calibration was
performed using measurements of the antenna gain and the system
temperature of each station.  Ionospheric corrections were applied
using the AIPS task ``TECOR''. The phase calibrators were
self-calibrated in both phase and amplitude, and imaged in an iterative
cycle. The self-calibration solutions of the phase calibrators were
applied to the respective target sources. To further improve the
signal-to-noise ratio of the images, we self calibrated the target sources themselves.

\section{RESULTS}

Figures 1, 2, and 3, are the VLBA images of J1053$-$0016, J1235$-$0003, and
J0913$+$5919, respectively. The rms noise in each image is
68.5, 67, and 66~$\mu$Jy~beam$^{-1}$, with dynamic ranges of 180, 252,
and 293, respectively. All these sources are unresolved at the angular
resolution of the VLBA at the observed 1.4~GHz frequency, suggesting that the radio
emission from each object is confined to a compact region less
than 10~mas in size.

Table 1 shows the results of fitting Gaussian models to the observed
source spatial profiles using the AIPS task ``JMFIT''.  The source
names, redshifts, and positions are listed in columns 1, 2,
3, and 4, respectively. Their total flux densities are listed in column 5, and the derived upper
limits to deconvolved source sizes are given in column 6.

The VLA flux densities of these objects at 1.4, 5, and 15 GHz, are
listed in Table~2 (Stern et al.2000; Carilli et al.~2001; Petric et al.~2003; M. P. Rupen
2003, private communication). In all cases, the 1.4 GHz flux densities
measured by the VLA and VLBA are equal to better than
$10\%$. These results imply that the sources are not highly variable
on time scales of years. Two of these objects, namely J1053$-$0016 and
J1235$-$0003, have flat spectra between 1.4 and 5~GHz, but have
steeper spectra between 5 and 15~GHz. The source J0913$+$5919 is a
steep spectrum compact source between 1.4 and 5~GHz. Note that an
observing frequency of 1.4 GHz corresponds to a rest frequency
of about 8 GHz at $z=4.5$.

We have synthesized larger images ($2'' \times 2''$) using the VLBA and found no
other radio components at $\ge 3\sigma$ level ($\sim200~\mu$Jy~beam$^{-1}$) in the field.
The implied dynamic range limit between the target sources themselves and any companion
structures is then larger than 60 to 100. A similar limit to extended and/or multiple structures
has been found on larger scales ($2''$ to $20''$) for all these three sources using the VLA
\citep{CAL01,PET03}.

\section{DISCUSSION}
The VLBA imaging of three $z> 4$ radio loud quasars (J1053$-$0016, J1235$-$0003, and
J0913$+$5919) shows that these sources are smaller than a few mas in size, corresponding to physical
scales of $\le 20$ pc. On the other hand, all three sources show
falling spectra at high frequency, and none of the sources are
variable on yearly timescales.  These results suggests that the
sources are likely core-jet radio sources, and hence may be used to
extend to very high redshifts the studies of cosmic geometry using
core-jet radio sources (Gurvits et al. 1999). Higher frequency VLBA
observations at higher spatial resolution are planned to test this
possibility.

We find no indication of multiple radio components in the fields of
these sources on scales of 20 mas to a few arcseconds, to
dynamic range limits of $\sim 100$.  A similar conclusion was reached
for three other $z>4$ radio loud quasars \citep{FRE97,FRE03}. These results imply that at least
these six high $z$ quasars are not strongly gravitationally lensed.

The compact nature of these quasars make them excellent
candidates for future H~{\footnotesize I} 21~cm absorption experiments to detect the
neutral IGM in their host galaxies \citep{FL02}, in
particular the steep spectrum source J0913$+$5919. Such a search is
currently under-way using the Giant Meter Wave Radio
Telescope. Knowledge of source structure, as presented herein, is
critical for both identifying potential candidates for H~{\footnotesize I} 21~cm
absorption searches, and for subsequent interpretation of the results.

\section{ACKNOWLEDGMENTS}

The authors thank G. B. Taylor for useful discussions.  This research
has made use of the NASA/IPAC Extragalactic Database (NED) which is
operated by the Jet Propulsion Laboratory, California Institute of
Technology, under contract with the National Aeronautics and Space
Administration.  E.~M. is grateful for support from NRAO through the
Pre-doctoral Research Program, and the NSF support through grant
AST~99-88341.

\clearpage
{\centerline{\large\bf Figure Captions}}

\figcaption[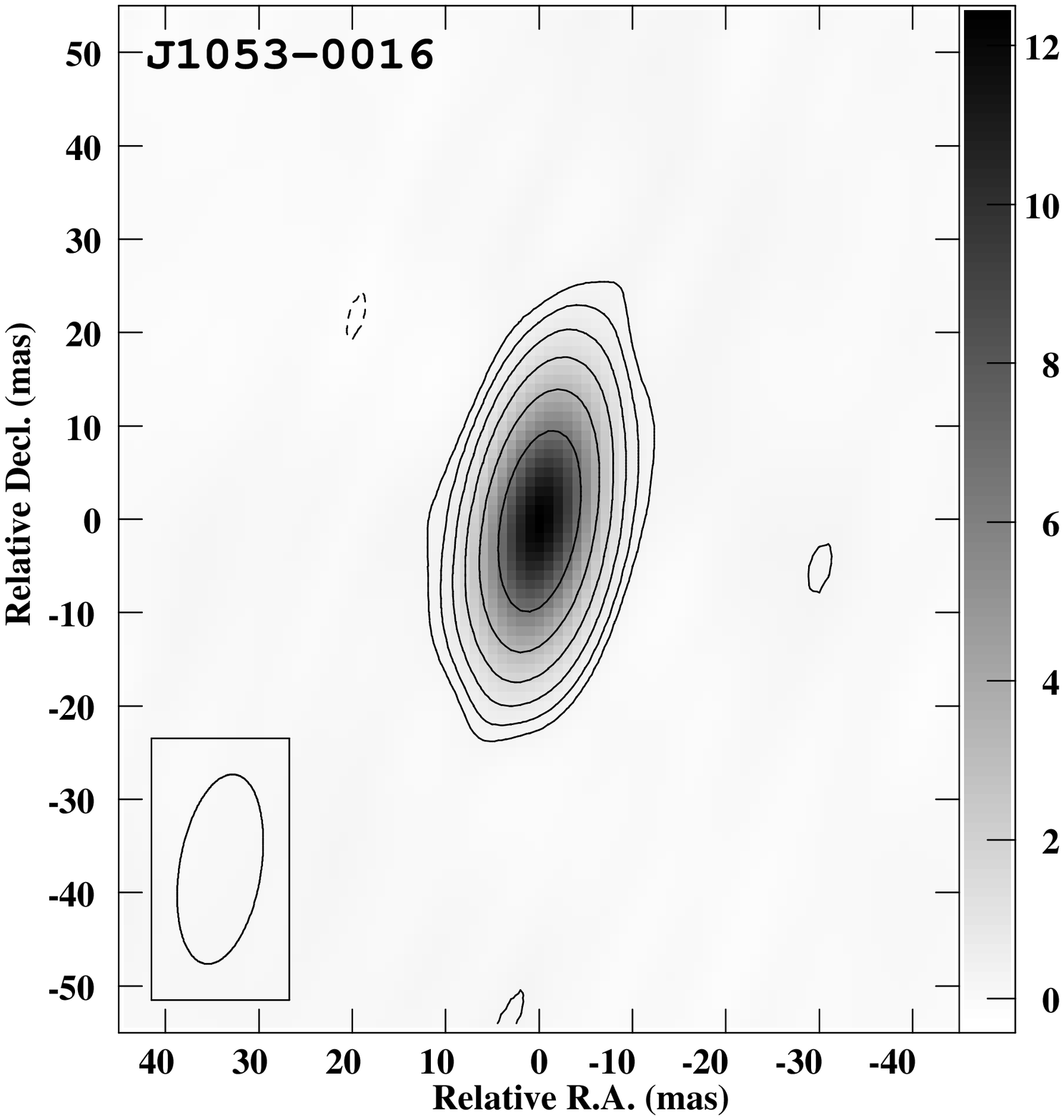]{Naturally weighted continuum image of
J1053$-$0016 ($z=4.29$) at 1.4~GHz. The restoring beam size is
$20.51 \times 8.74$~mas in position angle $-9^{\circ}$. The peak flux
density is 12.33~mJy~beam$^{-1}$, and the contour levels are at $-3$,
3, 6, 12, $\ldots$, 192 times the rms noise level, which is
68.5~$\mu$Jy~beam$^{-1}$. The gray-scale range is indicated by the
step wedge at the right side of the image. The reference point (0, 0)
is $\alpha(\rm{J2000.0})= 10^{\rm h}53^{\rm m}20\rlap{.}^{\rm s}$4264,
$\delta(\rm{J2000.0})=
-00^{\circ}16^{'}49\rlap{.}^{''}$6438.\label{f1}}

\figcaption[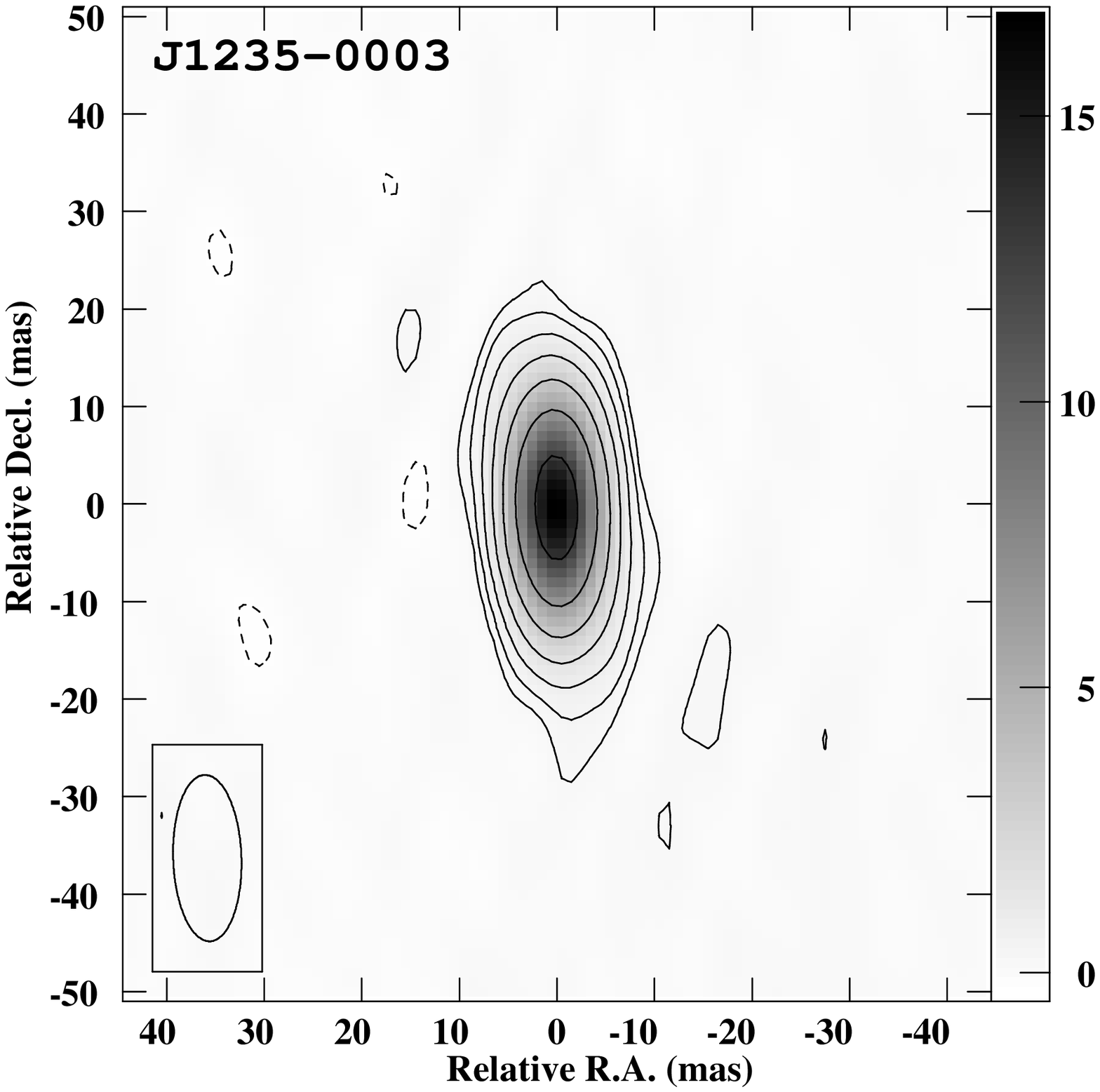]{Naturally weighted continuum image of
J1235$-$0003 ($z=4.69$) at 1.4~GHz. The restoring beam size is
$17.06 \times 6.99$~mas in position angle $2^{\circ}$. The peak flux
density is 16.91~mJy~beam$^{-1}$, and the contour levels are at $-3$,
3, 6, 12, $\ldots$, 192 times the rms noise level, which is
67~$\mu$Jy~beam$^{-1}$. The gray-scale range is indicated by the step
wedge at the right side of the image. The reference point (0, 0) is
$\alpha(\rm{J2000.0})= 12^{\rm h}35^{\rm m}03\rlap{.}^{\rm s}$0469,
$\delta(\rm{J2000.0})=
-00^{\circ}03^{'}31\rlap{.}^{''}$7606.\label{f2}}

\figcaption[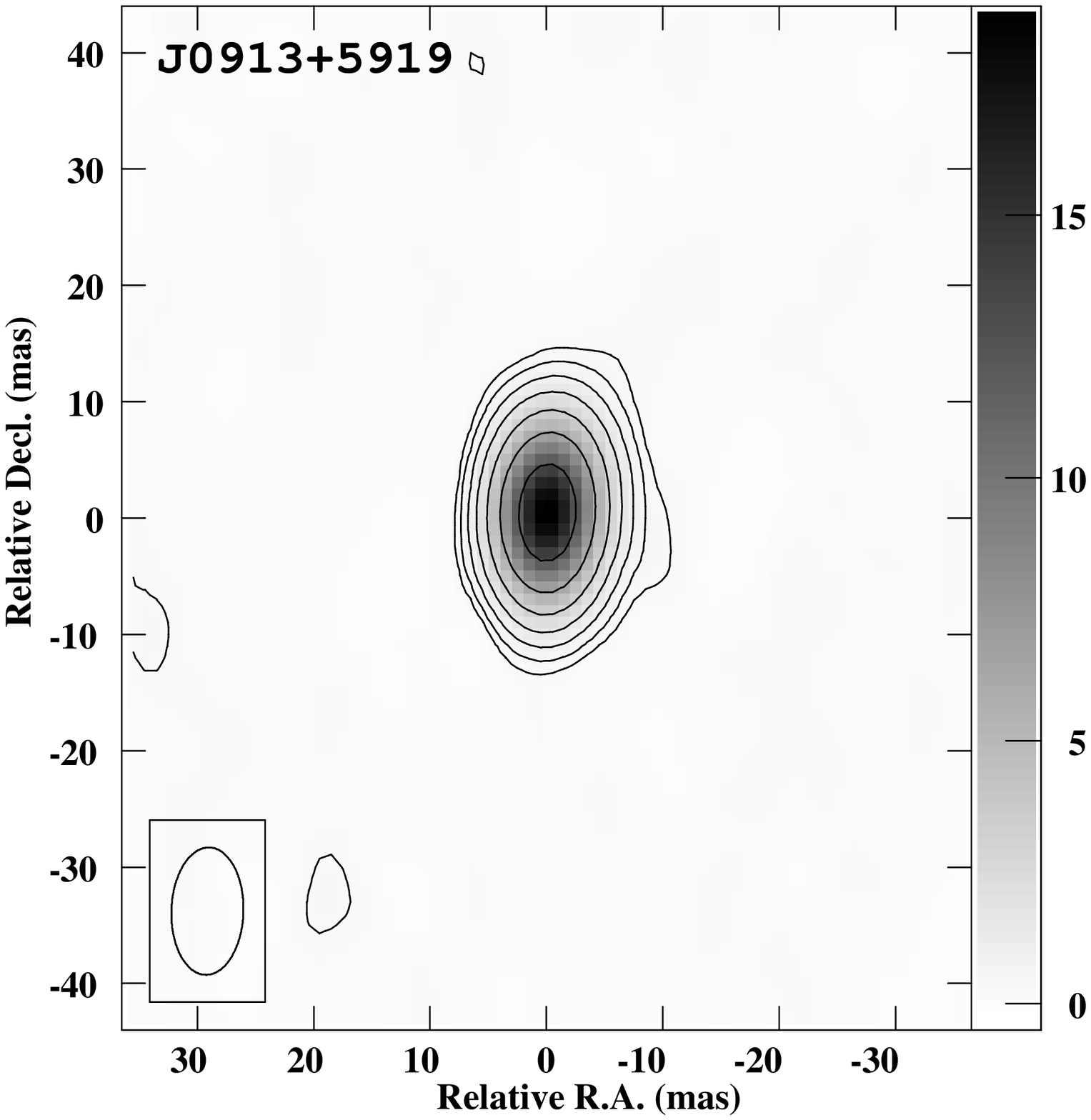]{Naturally weighted continuum image of
J0913+5919 ($z=5.11$) at 1.4~GHz. The restoring beam size is
$10.96 \times 6.19$~mas in position angle $-2^{\circ}$. The peak flux
density is 19.34~mJy~beam$^{-1}$, and the contour levels are at $-3$,
3, 6, 12, $\ldots$, 192 times the rms noise level, which is
66~$\mu$Jy~beam$^{-1}$. The gray-scale range is indicated by the step
wedge at the right side of the image. The reference point (0, 0) is
$\alpha(\rm{J2000.0})= 09^{\rm h}13^{\rm m}16\rlap{.}^{\rm s}$5472,
$\delta(\rm{J2000.0})=
+59^{\circ}19^{'}21\rlap{.}^{''}$6656.\label{f3}}

\clearpage
\begin{deluxetable}{cccccc}
\tablecolumns{9}
\tablewidth{0pc}
\tablecaption{S{\footnotesize OURCE} P{\footnotesize ARAMETERS}}
\tablehead{
\colhead{}    &\colhead{}
&\multicolumn{4}{c}{Gaussian Component Parameters} \\ \cline{3-6} \\
\colhead{Source} & \colhead{$z$} &\colhead{R.A. (J2000)}   &\colhead{Decl. (J2000)}&
\colhead{Total} & \colhead{Size} \\
\colhead{} & \colhead{}  &\colhead{(h m s)}   & \colhead{(${^\circ}~'~''$)}
& \colhead{(mJy)} & \colhead{(mas)} \\
\colhead{(1)}  & \colhead{(2)} & \colhead{(3)}   & \colhead{(4)} &
\colhead{(5)} & \colhead{(6)}}
\startdata
J1053$-$0016  & 4.29   &10 53 20.4264   &$-$00 16 49.6438 & $12.413 \pm  0.118$& $<3.2$ \\
J1235$-$0003  & 4.69   &12 35 03.0469   &$-$00 03 31.7606 & $17.155 \pm  0.116$& $<3.3$ \\
J0913$+$5919  & 5.11   &09 13 16.5472   &$+$59 19 21.6656 & $19.407 \pm  0.115$& $<1.3$ \\
\enddata
\end{deluxetable}

\clearpage
\begin{deluxetable}{cccccl}
\tablecolumns{10}
\tablewidth{0pc}
\tablecaption{VLA F{\footnotesize LUX} D{\footnotesize ENSITIES}}
\tablehead{
\colhead{Source} & & \colhead{$S_{1.4}$} & \colhead{$S_{5}$} & \colhead{$S_{15}$} &
\colhead{Ref.}\\
\colhead{} & & \colhead{(mJy)} &\colhead{(mJy)} &\colhead{(mJy)} \\
\colhead{(1)} & & \colhead{(2)}  & \colhead{(3)} & \colhead{(4)} & \colhead{(5)}}
\startdata
J1053$-$0016 &  & 11.5 & 10.2 & 6.5 &  1, 2, 3 \\
J1235$-$0003 &  & 18.8 & 17 & 6.9 &   1, 3\\
J0913$+$5919  &  & 18.95 & 8.1 & \nodata & 4\\
\enddata
\tablerefs{(1) Carilli et al. 2001; (2) Stern et al. 2000; (3) Petric et al. 2003; (4) M. P. Rupen
2003, private communication}
\end{deluxetable}

\clearpage
\begin{figure}
\epsscale{1}
\plotone{Momjian.fig1.eps}
\end{figure}

\clearpage
\begin{figure}
\epsscale{1}
\plotone{Momjian.fig2.eps}
\end{figure}

\clearpage
\begin{figure}
\epsscale{1}
\plotone{Momjian.fig3.eps}
\end{figure}

\end{document}